\documentclass[a4paper]{article}

\usepackage[margin=1in]{geometry}

\usepackage[T1]{fontenc}
\newcommand{\changefont}[3]{
\fontfamily{#1} \fontseries{#2} \fontshape{#3} \selectfont}

\changefont{ptm}{m}{n}

\usepackage{setspace} 
 \usepackage{graphicx}    

\newcommand \be{\begin{equation}}
\newcommand \ee{\end{equation}}
\newcommand \ba{\begin{eqnarray}}
\newcommand \ea{\end{eqnarray}}

\def\bit{\begin{itemize}}
\def\eit{\end{itemize}}

\usepackage{amsfonts}
\usepackage{amssymb}
\usepackage{amsmath}
\usepackage{graphics}
\usepackage{mathrsfs}
\usepackage{color}

\newtheorem{theorem}{Theorem}[section]

\newtheorem{lemma}{Lemma}[section]

\long\def\symbolfootnote[#1]#2{\begingroup%
\def\thefootnote{\fnsymbol{footnote}}\footnote[#1]{#2}\endgroup} 

\begin{document}

%

\begin{center}
\Large \textbf{Homoclinic and Heteroclinic Motions in Hybrid Systems with Impacts}
\end{center}

\vspace{-0.3cm}
\begin{center}
\normalsize \textbf{Mehmet Onur Fen$^{a,} \symbolfootnote[1]{Corresponding Author Tel.: +90 312 365 9276, E-mail: monur.fen@gmail.com}$, Fatma Tokmak Fen$^b$} \\
\vspace{0.2cm}
\textit{\textbf{\footnotesize$^a$Department of Mathematics, Middle East Technical University, 06800, Ankara, Turkey}} \\
\textit{\textbf{\footnotesize$^b$Department of Mathematics, Gazi University, 06500, Teknikokullar, Ankara, Turkey}}
\vspace{0.1cm}
\end{center}

\vspace{0.3cm}

\begin{center}
\textbf{Abstract}
\end{center}

\noindent\ignorespaces

In this paper, we present a method to generate homoclinic and heteroclinic motions in impulsive systems. We rigorously prove the presence of such motions in the case that the systems are under the influence of a discrete map that possesses homoclinic and heteroclinic orbits. Simulations that support the theoretical results are represented by means of a Duffing equation with impacts.

\vspace{0.2cm}
 
\noindent\ignorespaces \textbf{Keywords:} Impulsive systems; Stable and unstable sets; Homoclinic motion; Heteroclinic motion; Duffing equation with impacts

\section{Introduction}

Impulsive differential equations describe the dynamics of real world processes in which abrupt changes occur. Such equations play an increasingly important role in various fields such as mechanics, electronics, biology, neural networks, communication systems, chaos theory and population dynamics \cite{Akh1,Akh4,Akh3,Akh5,Herrera12,Khadra03,Liu94,Yang07,Yang97b,Zhou09}. In this paper, we investigate the existence of homoclinic and heteroclinic motions in systems with impulsive effects.

The main object of the present study is the following impulsive system,
\begin{eqnarray} \label{impulsive_system}
\begin{array}{l}
x'= A(t)x + f(t,x) + g(t,\zeta), ~ t\neq \theta_k, \\
\Delta x  |_{t= \theta_k} = B_k x + J_k (x) + \zeta_k,  
\end{array}
\end{eqnarray}
where $\left\{ \theta_k \right\},$ $k\in\mathbb Z,$ is a strictly increasing sequence of real numbers such that $\left|\theta_k\right| \to \infty$ as $\left| k \right| \to \infty,$ $A(t)$ is an $n \times n$ continuous matrix function, $B_k$ are constant $n \times n$  real valued matrices, $\Delta x  |_{t= \theta_k}=x(\theta_k+)-x(\theta_k),$ $x(\theta_k+)=\displaystyle \lim_{t\to\theta_k^+}x(t),$ the functions $f: \mathbb R \times \mathbb R^n \to \mathbb R^n$ and $J_k: \mathbb R^n \to \mathbb R^n$ are continuous in all their arguments, the function $g(t,\zeta)$ is defined by the equation $g(t,\zeta)=\zeta_k,$ $t \in (\theta_{k-1},\theta_k],$  and the sequence $\zeta=\left\{\zeta_k\right\},$ $k\in \mathbb Z,$ is a solution of the map 
\begin{eqnarray} \label{discrete_map}
\zeta_{k+1}=F(\zeta_k),
\end{eqnarray}
where the function $F:\Lambda \to \Lambda$ is continuous and $\Lambda$ is a bounded subset of $\mathbb R^n.$ Here, $\mathbb R$ and $\mathbb Z$ denote the sets of real numbers and integers, respectively. The system under investigation is a hybrid one, since it combines the dynamics of an impulsive differential equation with a discrete map. Our main objective is to prove rigorously the existence of homoclinic and heteroclinic solutions in the dynamics of (\ref{impulsive_system}) provided that (\ref{discrete_map}) possesses such solutions.

The idea of the usage of discontinuous perturbations to generate homoclinic and heteroclinic motions in systems of differential equations was first realized in the papers \cite{Akh2,Akh7} on the basis of functional spaces. It was shown in \cite{Akh2} that the chaotic attractor of the relay system, which was introduced in the paper \cite{Akh6}, consists of homoclinic solutions. Similar results for impulsive differential equations were obtained in the study \cite{Akh7} by taking advantage of the moments of impulses.

The existence of homoclinic and heteroclinic motions in systems with impulses were also investigated in the papers \cite{Battelli97,Fang12,Feckan96,Han11,Li14,Zhang14,Zhang11}. The existence and multiplicity of fast homoclinic solutions for a class of damped vibration problems with impulsive effects were investigated in \cite{Zhang14} by using the mountain pass theorem and the symmetric mountain pass theorem in the critical point theory. The mountain pass theorem was also utilized in \cite{Fang12,Li14} to show the presence of homoclinic motions in second order impulsive systems. On the other hand, Wei and Chen \cite{Wei14,Wei13} considered the existence of heteroclinic cycles in predator-prey systems with Allee effect and state-dependent impulsive harvesting within the scope of their studies.  Zhang and Li \cite{Zhang11} proved the existence of at least one non-zero homoclinic solution, which is generated by impulses, under appropriate conditions for a class of impulsive second order differential equations. Han and Zhang \cite{Han11} obtained the existence of homoclinic solutions for a class of asymptotically linear or sublinear Hamiltonian systems with impulses by using variational methods. It was mentioned in \cite{Han11} that no homoclinic solutions exist for the system under investigation without impulses. However, in the present study, the emergence of homoclinic and heteroclinic motions are completely provided by the influence of a discrete map instead of impulsive effects. Additionally, our results are valid for systems with arbitrary high dimensions.

The rest of the paper is organized as follows. In Section \ref{prelim}, we discuss bounded solutions of (\ref{impulsive_system}), and present sufficient conditions for the existence of homoclinic and heteroclinic motions in the system. Section \ref{mainresults} is devoted for the main results of the paper. In this part, we show the connection between the stable and unstable sets of the impulsive system (\ref{impulsive_system}) and the discrete map (\ref{discrete_map}), and prove the existence of homoclinic and heteroclinic solutions in (\ref{impulsive_system}). Examples concerning homoclinic and heteroclinic motions in an impulsive Duffing equation are provided in Section \ref{examples_sec}. Finally, some concluding remarks are given in Section \ref{conc}.

\section{Preliminaries} \label{prelim}

In the sequel, we will make use of the usual Euclidean norm for vectors and the norm induced by the Euclidean norm for matrices \cite{Horn85}.

Let us denote by $U(t,s)$ the transition matrix of the linear homogeneous system 
\begin{eqnarray} \label{homogeneous_system}
\begin{array}{l}
u'= A(t)u, ~ t\neq \theta_k, \\
\Delta u  |_{t= \theta_k} = B_k u(\theta_k).  
\end{array}
\end{eqnarray}

The following conditions are required.
\begin{enumerate}
\item[\textbf{(C1)}] $\det \left( I + B_k\right) \neq 0$ for all $k\in \mathbb Z,$ where $I$ is the $n\times n$ identity matrix;
\item[\textbf{(C2)}] There exists a positive number $\theta$ such that $\theta_{k+1}-\theta_k \ge \theta$ for all $k\in\mathbb Z;$
\item[\textbf{(C3)}] There exist positive numbers $N$ and $\omega$ such that $\left\| U(t,s)  \right\| \le Ne^{-\omega (t-s)}$ for $t\ge s;$ 
\item[\textbf{(C4)}] There exist positive numbers $M_f,$ $M_F$ and $M_J$ such that $$\displaystyle \sup_{(t,x)\in \mathbb R \times \mathbb R^n } \left\| f(t,x)\right\| \le M_f, \ \ \displaystyle \sup_{\sigma \in \Lambda} \left\| F(\sigma)\right\| \le M_F, \  \ \displaystyle \sup_{k\in\mathbb Z, x\in \mathbb R^n } \left\| J_k(x)\right\| \le M_J;$$ 
\item[\textbf{(C5)}] There exist positive numbers $L_f$ and $L_J$ such that $$\left\|f(t,x_1) - f(t,x_2)\right\| \le L_f \left\|x_1-x_2\right\|$$ for all $t\in \mathbb R,$ $x_1,x_2 \in \mathbb R^n,$ and  $$\left\|J_k(x_1) - J_k(x_2)\right\| \le L_J \left\|x_1-x_2\right\|$$ for all $k\in\mathbb Z,$ $x_1,x_2 \in \mathbb R^n;$
\item[\textbf{(C6)}] $\displaystyle N \left( \frac{L_f}{\omega} + \frac{ L_J}{1-e^{-\omega \theta}}    \right)<1;$ 
\item[\textbf{(C7)}] $-\omega+NL_f+\displaystyle\frac{1}{\theta}\ln(1+NL_J)<0.$
\end{enumerate}

Let $\Theta$ be the set of all sequences  $\zeta=\left\{\zeta_k\right\},$ $k\in \mathbb Z,$ obtained by equation (\ref{discrete_map}).  By using the results of \cite{Akh1,Samolienko95} one can show under the conditions $(C1)-(C6)$ that for a fixed sequence $\zeta \in \Theta$ the system (\ref{impulsive_system}) possesses a unique bounded on $\mathbb R$ solution $\phi_{\zeta}(t),$ which satisfies the following relation,
\begin{eqnarray} \label{bounded_soln_relation}
\phi_{\zeta} (t) = \displaystyle \int_{-\infty}^t U(t,s) \left[ f(s,\phi_{\zeta} (s)) + g(s,\zeta)  \right] ds 
+ \displaystyle \sum_{-\infty < \theta_k < t} U(t,\theta_k+) \left[J_k(\phi_{\zeta} (\theta_k))+\zeta_k\right].
\end{eqnarray} 
 
One can confirm under the conditions $(C1)-(C7)$ that for a fixed sequence $\zeta \in \Theta,$ the bounded solution $\phi_{\zeta}(t)$ attracts all other solutions of (\ref{impulsive_system}), i.e., $\left\|x(t)-\phi_{\zeta}(t)\right\|\to 0$ as $t\to \infty$ for any solution $x(t)$ of (\ref{impulsive_system}). Moreover, $$\displaystyle \sup_{t \in\mathbb R} \left\|\phi_{\zeta}(t)\right\| \le \displaystyle N \left(\displaystyle\frac{M_f+M_F}{\omega}+\displaystyle\frac{M_J+M_F}{1-e^{-\omega \theta}}\right)$$ for each $\zeta \in \Theta.$

\section{Homoclinic and heteroclinic motions} \label{mainresults}

In this section, first of all, we will describe the stable, unstable and hyperbolic sets as well as the homoclinic and heteroclinic motions for both system (\ref{impulsive_system}) and the discrete map (\ref{discrete_map}). These definitions were introduced in the papers \cite{Akh2,Akh7}. After that the existence of homoclinic and heteroclinic motions in the dynamics of (\ref{impulsive_system})  will be proved.

Consider the set $\Theta$ described in the previous section once again. The stable set of a sequence $\zeta\in\Theta$ is defined as 
\begin{eqnarray*} \label{stable_set}
W^s(\zeta)= \left\{ \eta \in \Theta \ | \ \left\|\eta_k-\zeta_k\right\|\to 0 ~\textrm{as}~ k\to  \infty  \right\},
\end{eqnarray*}
and the unstable set of $\zeta$ is 
\begin{eqnarray*} \label{unstable_set}
W^u(\zeta)= \left\{ \eta \in \Theta \ | \ \left\|\eta_k-\zeta_k\right\|\to 0 ~\textrm{as}~ k\to  -\infty  \right\}.
\end{eqnarray*}
The set $\Theta$ is called hyperbolic if for each $\zeta \in \Theta$ the stable and unstable sets of $\zeta$ contain at least one element different from $\zeta.$ A sequence $\eta \in \Theta$ is homoclinic to another sequence $\zeta \in \Theta$ if $\eta \in W^s(\zeta) \cap W^u(\zeta).$ Moreover, $\eta \in \Theta$ is heteroclinic to the sequences $\zeta^1 \in \Theta,$ $\zeta^2 \in \Theta,$ $\eta \neq \zeta^1,$ $\eta \neq \zeta^2,$ if  $\eta \in W^s(\zeta^1) \cap W^u(\zeta^2).$

On the other hand, let us denote by $\mathscr{A}$ the set consisting of all bounded on $\mathbb R$ solutions of system (\ref{impulsive_system}). 
A bounded solution $\phi_{\eta}(t) \in \mathscr{A}$ belongs to the stable set $W^s(\phi_{\zeta}(t))$ of  $\phi_{\zeta}(t) \in \mathscr{A}$ if $\left\|\phi_{\eta}(t)-\phi_{\zeta}(t)\right\|\to 0$ as $t\to  \infty.$ Besides,  $\phi_{\eta}(t)$ is an element of the unstable set $W^u(\phi_{\zeta}(t))$ of  $\phi_{\zeta}(t)$ provided that  $\left\|\phi_{\eta}(t)-\phi_{\zeta}(t)\right\|\to 0$ as $t\to -\infty.$ 

We say that $\mathscr{A}$ is  hyperbolic if for each  $\phi_{\zeta}(t) \in \mathscr{A}$ the sets $W^s(\phi_{\zeta}(t))$ and $W^u(\phi_{\zeta}(t))$ contain at least one element different from $\phi_{\zeta}(t).$ A solution $\phi_{\eta}(t)\in \mathscr{A}$ is homoclinic to another solution $\phi_{\zeta}(t) \in \mathscr{A}$ if $\phi_{\eta}(t) \in W^s(\phi_{\zeta}(t)) \cap W^u(\phi_{\zeta}(t)),$ and 
$\phi_{\eta}(t)\in \mathscr{A}$ is heteroclinic to the bounded solutions $\phi_{\zeta^1}(t),$ $\phi_{\zeta^2}(t)\in \mathscr{A},$ $\phi_{\eta}(t) \neq \phi_{\zeta^1}(t),$ $\phi_{\eta}(t) \neq \phi_{\zeta^2}(t),$ if $\phi_{\eta}(t) \in W^s(\phi_{\zeta^1}(t)) \cap W^u(\phi_{\zeta^2}(t)).$

In what follows, we will denote by $i((a,b))$ the number of the terms of the sequence $\left\{\theta_k\right\},$ $k\in \mathbb Z,$ which belong to the interval $(a,b),$ where $a$ and $b$ are real numbers such that $a<b.$ It is worth noting that $\displaystyle i((a,b))\leq 1+\frac{b-a}{\theta}.$

The connection between the stable sets of the solutions of (\ref{impulsive_system}) and (\ref{discrete_map}) is provided in the next assertion.

\begin{lemma}\label{lemma1} 
Suppose that the conditions $(C1)-(C7)$ are fulfilled, and let $\zeta$ and $\eta$ be elements of $\Theta.$ If $\eta\in W^{s}(\zeta),$ then $\phi_{\eta}(t)\in W^{s}(\phi_{\zeta}(t)).$  
\end{lemma}

\noindent \textbf{Proof.} 
Fix an arbitrary positive number $\epsilon,$ and denote $\alpha= \omega-NL_f-\displaystyle\frac{1}{\theta}\ln(1+NL_J).$ Assume without loss of generality that $\epsilon \le 2 M_F.$ Let $\gamma$ be a real number such that  
$$\gamma \ge 1+N\left(\frac{1}{\omega}+\frac{1}{1-e^{-\omega\theta}}\right)\left(1+\frac{N L_f(1+N L_J)}{\alpha}+\frac{N L_J(1+N L_J)}{1-e^{-\alpha \theta}}\right).$$

Because the sequence $\eta=\left\{\eta_k\right\},$ $k\in\mathbb Z,$ belongs to the stable set $W^{s}(\zeta)$ of $\zeta=\left\{\zeta_k\right\},$ there exists an integer $k_0$ such that $\left\|\eta_k-\zeta_k\right\|<\displaystyle\frac{\epsilon}{\gamma}$ for all $k\ge k_0.$ One can confirm that $\left\|g(t,\eta)-g(t,\zeta)\right\|<\displaystyle \frac{\epsilon}{\gamma}$ for $t>\theta_{k_0-1}.$

Making use of the relation
\begin{eqnarray*} 
&& \phi_{\eta}(t) - \phi_{\zeta}(t) = \displaystyle \int_{-\infty}^{t} U(t,s) \left[ f(s,\phi_{\eta}(s)) -  f(s,\phi_{\zeta}(s))  + g(s,\eta)- g(s,\zeta) \right] ds \\
&& + \displaystyle \sum_{-\infty < \theta_k < t} U(t,\theta_k+) \left[ J_k (\phi_{\eta}(\theta_k)) - J_k (\phi_{\zeta}(\theta_k)) +\eta_k - \zeta_k \right],
\end{eqnarray*}
we obtain for $t>\theta_{k_0-1}$ that
\begin{eqnarray} \label{proof_ineq1}
\begin{array}{l}
\displaystyle  \left\|\phi_{\eta}(t) - \phi_{\zeta}(t)\right\| \le \displaystyle \int_{-\infty}^{\theta_{k_0-1}} 2N(M_f+M_F) e^{-\omega (t-s)} ds  \\ 
+ \displaystyle \sum_{-\infty < \theta_k \le \theta_{k_0-1}} 2N(M_J+M_F) e^{-\omega (t-\theta_k)} + \displaystyle \int^{t}_{\theta_{k_0-1}} \frac{N\epsilon}{\gamma} e^{-\omega (t-s)} ds\\
 + \displaystyle  \sum_{\theta_{k_0-1} < \theta_k < t } \frac{N\epsilon}{\gamma} e^{-\omega (t-\theta_k)} + \displaystyle \int^{t}_{\theta_{k_0-1}} NL_f  e^{-\omega (t-s)} \left\|\phi_{\eta}(s) - \phi_{\zeta}(s)\right\| ds \\
+\displaystyle  \sum_{\theta_{k_0-1} < \theta_k < t } NL_J e^{-\omega (t-\theta_k)} \left\|\phi_{\eta}(\theta_k) - \phi_{\zeta}(\theta_k)\right\| \\
\le \displaystyle  2N \left(   \frac{  M_f+M_F }{\omega}  +    \frac{ M_J+M_F }{1-e^{-\omega \theta} }   \right)e^{-\omega (t-\theta_{k_0-1})} \\
 + \displaystyle \frac{N \epsilon}{\gamma \omega} \left( 1-e^{-\omega(t-\theta_{k_0-1})}  \right) + \displaystyle \frac{N \epsilon}{\gamma (1-e^{-\omega \theta})} \left( 1-e^{-\omega(t-\theta_{k_0-1} + \theta)}  \right) \\
+ \displaystyle \int^{t}_{\theta_{k_0-1}} NL_f  e^{-\omega (t-s)} \left\|\phi_{\eta}(s) - \phi_{\zeta}(s)\right\| ds \\
+\displaystyle  \sum_{\theta_{k_0-1} < \theta_k < t } NL_J e^{-\omega (t-\theta_k)} \left\|\phi_{\eta}(\theta_k) - \phi_{\zeta}(\theta_k)\right\|. 
\end{array}
\end{eqnarray}

Define the functions $u(t)=e^{\omega t} \left\|  \phi_{\eta}(t) - \phi_{\zeta}(t)  \right\|$ and $h(t)= c_1 + c_2 e^{\omega t},$ where
$$c_1=2N \left(   \frac{  M_f+M_F }{\omega}  +    \frac{ M_J+M_F }{1-e^{-\omega \theta} }   \right)e^{\omega \theta_{k_0-1}} - \frac{N\epsilon}{\gamma} \left(\frac{e^{\omega \theta_{k_0-1}}}{\omega}+ \frac{e^{\omega (\theta_{k_0-1}-\theta)}}{1-e^{-\omega \theta}}\right)$$ and $$c_2=\frac{N\epsilon}{\gamma} \left(\frac{1}{\omega}+ \frac{1}{1-e^{-\omega \theta}}\right).$$
The inequality (\ref{proof_ineq1}) implies that 
$$
u(t) \le h(t) + \displaystyle \int_{\theta_{k_0-1}}^t  NL_f u(s) ds + \sum_{\theta_{k_0-1} < \theta_k < t} NL_J u(\theta_k).
$$
The application of the analogue of the Gronwall's inequality for piecewise continuous functions yields
\begin{eqnarray*}
&& u(t) \le h(t) + \displaystyle \int_{\theta_{k_0-1}}^t  NL_f (1+ NL_J)^{i((s,t))} e^{NL_f(t-s)} h(s) ds  \\
&&+ \displaystyle \sum_{\theta_{k_0-1} < \theta_k < t} NL_J (1+NL_J)^{i((\theta_k,t))} e^{NL_f(t-\theta_k)} h(\theta_k).
\end{eqnarray*}
Since the equation
\begin{eqnarray*}
&& 1+\displaystyle \int^t_{\theta_{k_0-1}} NL_f (1+NL_J)^{i((s,t))} e^{NL_f(t-s)} ds \\
&& + \displaystyle \sum_{\theta_{k_0-1} < \theta_k < t} NL_J (1+NL_J)^{i((\theta_k,t))} e^{NL_f(t-\theta_k)} \\
&& = (1+ NL_J)^{i((\theta_{k_0-1},t))} e^{NL_f(t-\theta_{k_0-1})}
\end{eqnarray*}
is valid and $(1+NL_J)^{i((a,b))} e^{NL_f(b-a)} \le (1+NL_J) e^{(\omega-\alpha)(b-a)} $ for any real numbers $a$ and $b$ with $a < b,$ one can confirm that
\begin{eqnarray*}
&& u(t) \le c_1 (1+NL_J) e^{(\omega-\alpha)(t-\theta_{k_0-1})} + c_2e^{\omega t}  \\
&& + \displaystyle \int_{\theta_{k_0-1}}^t c_2NL_f (1+NL_J) e^{(\omega-\alpha) (t-s)} e^{\omega s} ds \\
&& + \displaystyle \sum_{\theta_{k_0-1}<\theta_k< t} c_2 NL_J (1+NL_J) e^{(\omega-\alpha)(t-\theta_k)} e^{\omega \theta_k} \\
&& \le c_1 (1+NL_J) e^{(\omega-\alpha)(t-\theta_{k_0-1})} + c_2e^{\omega t}  \\
&& + \displaystyle \frac{c_2 NL_f (1+NL_J)}{\alpha} e^{\omega t} \left( 1-e^{-\alpha (t-\theta_{k_0-1})}  \right) \\
&& +  \displaystyle \frac{c_2 NL_J (1+NL_J)}{1-e^{-\alpha \theta}} e^{\omega t} \left( 1-e^{-\alpha (t-\theta_{k_0-1}+\theta)}  \right).
\end{eqnarray*}
If we multiply both sides of the last inequality by $e^{-\omega t},$ then we obtain that  
\begin{eqnarray*}
&& \left\|\phi_{\eta}(t) - \phi_{\zeta}(t) \right\| \le c_1 (1+NL_J) e^{-\omega \theta_{k_0-1}} e^{-\alpha(t-\theta_{k_0-1})} + c_2\\
&& + \displaystyle \frac{c_2 NL_f (1+NL_J)}{\alpha}   \left( 1-e^{-\alpha (t-\theta_{k_0-1})}  \right) \\
&& + \displaystyle \frac{c_2 NL_J (1+NL_J)}{1-e^{-\alpha \theta}}  \left( 1-e^{-\alpha (t-\theta_{k_0-1}+\theta)}  \right) \\
&& < 2N(1+NL_J) \left( \frac{M_f+M_F}{\omega} + \frac{M_J+M_F}{1-e^{-\omega \theta}}  \right) e^{-\alpha (t-\theta_{k_0-1})} \\
&& + \displaystyle \frac{N\epsilon}{\gamma}\left(\frac{1}{\omega}+\frac{1}{1-e^{-\omega\theta}}\right)\left(1+\frac{N L_f(1+N L_J)}{\alpha}+\frac{N L_J(1+N L_J)}{1-e^{-\alpha \theta}}\right).
\end{eqnarray*}

Now, let $R> \theta_{k_0-1}$ be a sufficiently large real number such that
\begin{eqnarray*}
\displaystyle 2N(1+N L_J)\left(\displaystyle\frac{M_f+M_F}{\omega}+\frac{M_J+M_F}{1-e^{-\omega \theta}}\right)e^{-\alpha (R-\theta_{k_0-1})}\le \frac{\epsilon}{\gamma}.
\end{eqnarray*}
For $t\ge R,$ we have
\begin{eqnarray*}
 \Big\|\phi_{\eta}(t)-\phi_{\zeta}(t)\Big\|<\displaystyle\frac{\epsilon}{\gamma}\Big[1+N\Big(\frac{1}{\omega}+\frac{1}{1-e^{-\omega\theta}}\Big) 
 \Big(1+\frac{N L_f(1+N L_J)}{\alpha}+\frac{N L_J(1+N L_J)}{1-e^{-\alpha \theta}}\Big)\Big] 
 \le \epsilon.
\end{eqnarray*}
Therefore, $\displaystyle \lim_{t \to \infty} \left\|\phi_{\eta}(t)-\phi_{\zeta}(t)\right\|=0.$ Consequently, $\phi_{\eta}(t)\in W^{s}(\phi_{\zeta}(t)).$ $\square$

In the next lemma, we reveal the connection between the unstable sets of the solutions of (\ref{impulsive_system}) and (\ref{discrete_map}). 

\begin{lemma}\label{lemma2} 
Suppose that the conditions $(C1)-(C6)$ are fulfilled, and let $\zeta$ and $\eta$ be elements of $\Theta.$ If  $\eta\in W^{u}(\zeta),$ then $\phi_{\eta}(t)\in W^{u}(\phi_{\zeta}(t)).$ 
\end{lemma}

\noindent \textbf{Proof.} 
Fix an arbitrary positive number $\epsilon,$ and let $\lambda$ be a real number such that  
$$\lambda > \frac{N(\omega + 1 -e^{-\omega\theta})}{\omega(1-e^{-\omega\theta})-N(L_f(1-e^{-\omega\theta})+L_J\omega)}.$$

Since $\eta=\left\{\eta_k\right\},$ $k\in\mathbb Z,$ is an element of the unstable set $W^{u}(\zeta)$ of $\zeta=\left\{\zeta_k\right\},$ there exists an integer $k_0$ such that $\left\|\eta_k-\zeta_k\right\|<\displaystyle\frac{\epsilon}{\lambda}$ for all $k\le k_0.$ In this case, we have that $\left\|g(t,\eta)-g(t,\zeta)\right\|<\displaystyle \frac{\epsilon}{\lambda}$ for $t\le\theta_{k_0}.$

By using the relation
\begin{eqnarray*} 
&& \phi_{\eta}(t) - \phi_{\zeta}(t) = \displaystyle \int_{-\infty}^{t} U(t,s) \left[ f(s,\phi_{\eta}(s)) -  f(s,\phi_{\zeta}(s))  + g(s,\eta)- g(s,\zeta) \right] ds \\
&& + \displaystyle \sum_{-\infty < \theta_k < t} U(t,\theta_k+) \left[ J_k (\phi_{\eta}(\theta_k)) - J_k (\phi_{\zeta}(\theta_k)) +\eta_k - \zeta_k \right],
\end{eqnarray*}
one can verify for $t \le \theta_{k_0}$ that 
\begin{eqnarray*} 
&& \left\|\phi_{\eta}(t) - \phi_{\zeta}(t)\right\| < \displaystyle \int^t_{-\infty} N e^{-\omega (t-s)} \left( L_f  \left\|\phi_{\eta}(s) - \phi_{\zeta}(s)\right\| +\frac{\epsilon}{\lambda} \right) ds \\
&& + \displaystyle \sum_{-\infty < \theta_k < t} N e^{-\omega (t-\theta_k)} \left( L_J  \left\|\phi_{\eta}(\theta_k) - \phi_{\zeta}(\theta_k)\right\| +\frac{\epsilon}{\lambda} \right) \\
&& \le \frac{N}{\omega} \left( L_f \sup_{t\le\theta_{k_0}}   \left\|  \phi_{\eta}(t) - \phi_{\zeta}(t)   \right\|+ \frac{\epsilon}{\lambda}  \right) 
 + \frac{N}{1-e^{-\omega \theta}} \left( L_J \sup_{t\le\theta_{k_0}}   \left\|  \phi_{\eta}(t) - \phi_{\zeta}(t)  \right\|  + \frac{\epsilon}{\lambda} \right).
\end{eqnarray*}
Therefore,
\begin{eqnarray*} 
\left( 1-\frac{NL_f}{\omega}-\frac{NL_J}{1-e^{-\omega \theta}}  \right)    \sup_{t\le\theta_{k_0}}   \left\|  \phi_{\eta}(t) - \phi_{\zeta}(t)  \right\|  \le \frac{N\epsilon}{\lambda} \left( \frac{1}{\omega} + \frac{1}{1-e^{-\omega \theta}}  \right).
\end{eqnarray*}
The last inequality implies that $\displaystyle\sup_{t\le\theta_{k_0}}   \left\|  \phi_{\eta}(t) - \phi_{\zeta}(t)  \right\|  < \epsilon.$ Consequently, $$\displaystyle \lim_{t \to -\infty} \left\|\phi_{\eta}(t)-\phi_{\zeta}(t)\right\|=0,$$ and $\phi_{\eta}(t)$ belongs to $W^u(\phi_{\zeta}(t)).$ $\square$

The main result of the present paper is mentioned in the following theorem, which can be proved by using the results of Lemma \ref{lemma1} and Lemma \ref{lemma2}.

\begin{theorem}\label{main_theorem}
Under the conditions $(C1)-(C7),$ the following assertions are valid.
\begin{enumerate}
\item[(i)] If $\eta \in \Theta$ is homoclinic to $\zeta \in \Theta,$ then $\phi_{\eta}(t) \in \mathscr{A}$ is homoclinic to $\phi_{\zeta}(t) \in \mathscr{A};$
\item[(ii)] If $\eta \in \Theta$ is heteroclinic to $\zeta^1,$ $\zeta^2 \in \Theta,$ then $\phi_{\eta}(t) \in \mathscr{A}$ is heteroclinic  to $\phi_{\zeta^1}(t),$ $\phi_{\zeta^2}(t) \in \mathscr{A};$
\item[(iii)] If $\Theta$ is hyperbolic, then the same is true for $\mathscr{A}.$
\end{enumerate}
\end{theorem}

The next section is devoted to examples concerning homoclinic and heteroclinic motions in an impulsive Duffing equation.

\section{Examples} \label{examples_sec}

Let us take into account the impulsive Duffing equation
\begin{eqnarray} \label{imp_Duf}
\begin{array}{l}
x'' + 0.2 x' + 0.81 x + 0.001 x^3 = 0.7 \displaystyle \cos\left(\frac{2\pi}{3} t\right) + g(t,\zeta), \ t\neq \theta_k, \\
\Delta x  |_{t= \theta_k} =  -0.12 x  + 0.09+\zeta_k,  \\
\Delta x'  |_{t= \theta_k} =  -0.12 x'+ 0.015 \sin(x), 
\end{array}
\end{eqnarray}
where $\theta_k=3k,$ $k\in \mathbb Z,$ the function $g(t,\zeta)$ is defined through the equation $g(t,\zeta)=\zeta_k,$ $t \in (\theta_{k-1},\theta_k],$  and the sequence $\zeta=\left\{\zeta_k\right\}$ is a solution of the logistic map 
\begin{eqnarray} \label{logistic_map}
\zeta_{k+1}=F_{\mu}(\zeta_k),
\end{eqnarray}
where $F_{\mu}(s)=\mu s (1-s)$ and $\mu$ is a parameter. 

For $0<\mu\leq 4,$ the interval $[0,1]$ is invariant under the iterations of (\ref{logistic_map}) \cite{Dev90,Hale91,Rob95}, and the inverses of the function $F_{\mu}$ on the intervals $[0,1/2]$ and $[1/2,1]$ are
$
h_1(s)=\displaystyle \frac{1}{2} \left( 1-\sqrt{1-\frac{4s}{\mu}} \right)
$
and
$
h_2(s)=\displaystyle \frac{1}{2} \left( 1+\sqrt{1-\frac{4s}{\mu}} \right),
$
respectively.

By using the new variables $x_1=x$ and $x_2=x'$ one can reduce (\ref{imp_Duf}) to the system
\begin{eqnarray} \label{imp_Duf_system}
\begin{array}{l}
x_1'=x_2, \\
x_2'  = - 0.81 x_1 - 0.2 x_2 - 0.001 x_1^3 + 0.7 \displaystyle \cos \left(\frac{2\pi}{3} t\right) + g(t,\zeta), \ t\neq \theta_k, \\
\Delta x_1  |_{t= \theta_k} =  -0.12 x_1 +0.09+ \zeta_k,  \\
\Delta x_2  |_{t= \theta_k} =  -0.12 x_2+ 0.015 \sin(x_1) . 
\end{array}
\end{eqnarray}
Denote by $U(t,s)$ the transition matrix of the linear homogeneous system 
\begin{eqnarray} \label{linear_homogenous_system_imp}
\begin{array}{l}
u'_1=u_2, \\
u'_2=-0.81 u_1 - 0.2 u_2, ~t \neq \theta_k, \\
\Delta u_1|_{t=\theta_k} = -0.12 u_1, \\
\Delta u_2|_{t=\theta_k} = -0.12 u_2.
\end{array}
\end{eqnarray}
One can verify for $t> s$ that 
\[U(t,s)= e^{-(t-s)/10}   \left(\frac{22}{25}\right)^{i([s,t))}  P\left(
\begin {array}{ccc}
  \cos \Big(\frac{2}{\sqrt{5}}(t-s)\Big)&- \sin \Big(\frac{2}{\sqrt{5}}(t-s)\Big)\\
\noalign{\medskip}
 \sin \Big(\frac{2}{\sqrt{5}}(t-s)\Big)& \cos \Big(\frac{2}{\sqrt{5}}(t-s)\Big)
\end {array}
\right)P^{-1},\]
where $i([s,t))$ is the number of the terms of the sequence $\left\{\theta_k\right\}$ that belong to the interval $[s,t)$ and 
$P=\left(
\begin {array}{ccc}
0&1\\
\noalign{\medskip}
2/\sqrt{5}&-1/10
\end {array}
\right).$ 
It can be calculated that $\left\|U(t,s)\right\|\le N e^{-\omega (t-s)},$ $t\ge s,$ where $\omega=1/10$ and $N=1.17.$

For $0 < \mu \le 4$ the bounded solutions of (\ref{imp_Duf_system}) lie inside the compact region $$D=\left\{ (x_1,x_2) \in \mathbb R^2: \left|x_1\right| \le 2.8, \  \left|x_2\right| \le 1.4 \right\},$$ and the conditions $(C1)-(C7)$ are valid for system (\ref{imp_Duf_system}). It is worth noting that for a periodic solution $\zeta=\left\{\zeta_k\right\}$ of (\ref{logistic_map}) the corresponding bounded solution $\phi_{\zeta}(t)$ of (\ref{imp_Duf_system}) is also periodic.

\begin{figure}[ht] 
\centering
\includegraphics[width=11.0cm]{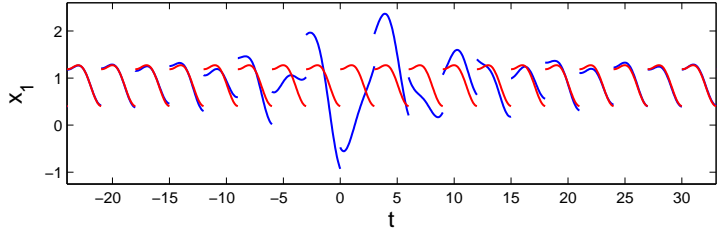}
\caption{\footnotesize Homoclinic solution of (\ref{imp_Duf_system}). The $x_1-$coordinates of $\phi_{\eta}(t)$ and $\phi_{\eta^*}(t)$ are shown in blue and red colors, respectively. The figure manifests that $\phi_{\eta}(t)$ is homoclinic to $\phi_{\eta^*}(t).$ }
\label{fig1}
\end{figure}

Consider the map (\ref{logistic_map}) with $\mu=3.9.$ It was demonstrated in \cite{Avrutin15} that the orbit $$\eta=\left\{\ldots, h^3_2(\eta_0), h_2^2(\eta_0), h_2(\eta_0), \eta_0, F_{\mu}(\eta_0), F^2_{\mu}(\eta_0), F^3_{\mu}(\eta_0), \ldots \right\},$$ where $\eta_0=1/3.9,$ is homoclinic to the fixed point $\eta^{*}=2.9/3.9$ of (\ref{logistic_map}). Denote by $\phi_{\eta}(t)$ and $\phi_{\eta^*}(t)$ the bounded solutions of (\ref{imp_Duf_system}) corresponding to $\eta$ and $\eta^*,$ respectively. One can conclude by using Theorem \ref{main_theorem}  that $\phi_{\eta}(t)$ is homoclinic to the periodic solution $\phi_{\eta^*}(t).$ Figure \ref{fig1} shows the graphs of the $x_1-$coordinates of $\phi_{\eta}(t)$ and $\phi_{\eta^*}(t).$ In the figure, the solution $\phi_{\eta}(t)$ is represented in blue color, while $\phi_{\eta^*}(t)$ is represented in red color. Figure \ref{fig1} reveals that $\phi_{\eta}(t)$ is homoclinic to $\phi_{\eta^*}(t),$ i.e., $\left\|\phi_{\eta}(t)-\phi_{\eta^*}(t)\right\| \to 0$ as $t \to \pm \infty.$

\begin{figure}[ht] 
\centering
\includegraphics[width=11.0cm]{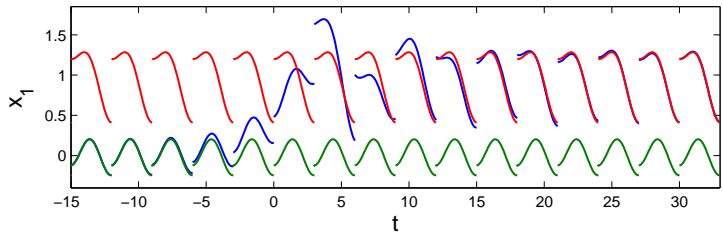}
\caption{\footnotesize Heteroclinic solution of (\ref{imp_Duf_system}). The $x_1-$coordinates of $\phi_{\widetilde{\eta}}(t),$ $\phi_{\eta^1}(t)$ and $\phi_{\eta^2}(t)$ are represented in blue, red and green colors, respectively. The figure confirms that $\phi_{\widetilde{\eta}}(t)$ is heteroclinic to the periodic solutions $\phi_{\eta^1}(t),$ $\phi_{\eta^2}(t).$}
\label{fig2}
\end{figure}

Now, we set $\mu=4$ in equation (\ref{logistic_map}). According to \cite{Avrutin15}, the orbit $$\widetilde{\eta}=\left\{\ldots, h^3_1(\widetilde{\eta}_0), h_1^2(\widetilde{\eta}_0), h_1(\widetilde{\eta}_0), \widetilde{\eta}_0, F_{\mu}(\widetilde{\eta}_0), F^2_{\mu}(\widetilde{\eta}_0), F^3_{\mu}(\widetilde{\eta}_0), \ldots \right\},$$ where $\widetilde{\eta}_0=1/4,$ is heteroclinic to the fixed points $\eta^1=3/4$ and $\eta^2=0$ of (\ref{logistic_map}). Suppose that $\phi_{\widetilde{\eta}}(t),$ $\phi_{\eta^1}(t)$ and $\phi_{\eta^2}(t)$ are the bounded solutions of (\ref{imp_Duf_system}) corresponding to $\widetilde{\eta},$ $\eta^1$ and $\eta^2,$ respectively. Theorem \ref{main_theorem} implies that $\phi_{\widetilde{\eta}}(t)$ is heteroclinic to the periodic solutions $\phi_{\eta^1}$ and $\phi_{\eta^2}.$ Figure \ref{fig2} shows the graphs of the $x_1-$coordinates of $\phi_{\widetilde{\eta}}(t),$ $\phi_{\eta^1}(t)$ and $\phi_{\eta^2}(t)$ in blue, red and green colors, respectively. The figure supports Theorem \ref{main_theorem} such that $\phi_{\widetilde{\eta}}(t)$ converges to $\phi_{\eta^1}(t)$ as time increases and converges to $\phi_{\eta^2}(t)$ as time decreases, i.e., $\phi_{\widetilde{\eta}}(t)$ is heteroclinic to $\phi_{\eta^1}(t),$ $\phi_{\eta^2}(t).$

\section{Conclusions} \label{conc}

In this study, we rigorously prove the presence of homoclinic and heteroclinic motions in hybrid systems with impacts. The dynamics of the system under consideration consist of an impulsive differential equation and a discrete map, which influences the former. According to our results, homoclinic and heteroclinic orbits of the discrete map give rise to the emergence of homoclinic and heteroclinic motions in the impulsive system. The presented technique is appropriate to design mechanical and electrical impulsive systems with homoclinic and heteroclinic motions, without any restriction in the dimension. One can take advantage of our approach to investigate the presence of such motions in hybrid systems with impacts. An impulsive Duffing equation is utilized to illustrate the results of the paper. The provided examples show the applicability of our results.

\section*{Acknowledgments}

The authors wish to express their sincere gratitude to the referees for the helpful criticism and valuable suggestions, which helped to improve the paper significantly. 

This work is supported by the 2219 scholarship programme of T\"{U}B\.{I}TAK, the Scientific and Technological Research Council of Turkey.

\end{document}